\documentstyle[twocolumn,prl,aps,epsfig]{revtex}
\title{Two-Dimensional Copolymers and Exact Conformal Multifractality}
\begin{document}
\input epsf
\input mssymb
\draft
\author{Bertrand Duplantier}
\address{Service de Physique Th\'eorique de Saclay, F-91191
Gif-sur-Yvette Cedex}
\address{and Institut Henri Poincar\'e, 11, rue
Pierre et Marie Curie, F-75231 Paris Cedex 05}
\date{August 21, 1998}
\maketitle
\begin{abstract}
{\bf Abstract.}
We consider in two dimensions (2D) the most general star-shaped copolymer,
mixing random (RW) or self-avoiding walks (SAW) with specific mutual
avoidance interactions thereof. Its exact conformal scaling dimensions in the
plane are derived from an algebraic structure existing 
on a random lattice (2D quantum gravity). 
The 
multifractal dimensions $\tau (n)$ of the harmonic measure of a
2D RW or SAW are conformal dimensions of certain star
copolymers. The exact associated $f(\alpha )$ are identical
for a RW or a SAW in 2D. These are examples of
conformal multifractality.
\end{abstract}
The concepts of generalized dimensions and associated multifractal (Mf)
measures have been developed more than a decade ago \cite
{mandelbrot,hentschel,frisch,halsey}. They are encountered in many physical
situations: strange attractors in dynamical systems, growth phenomena,
harmonic measure of diffusion-limited aggregates, electron localization,
random resistors, and random spin systems. Recently, analytic progress has
been made for turbulence of a passive scalar \cite{bernard} or for a growth
model of DLA \cite{halsey3}, both where the Mf dimensions can be calculated
perturbatively.

It is well known that universal geometrical fractals, e.g., random walks,
polymers, Ising or percolation models are essentially related to standard
critical phenomena and field theory, for which conformal invariance in two
dimensions (2D) has brought a wealth of exact results. By contrast, few
connections between multifractals and field theory have been found, although
the algebras of their respective correlation functions reveal intriguing
similarities \cite{cates}. It remains a challenge to see if an exact
description of some multifractal phenomena could emerge in 2D from the
conformal invariance classification.

A particularly interesting multifractal phenomenon was singled out some time
ago by Cates and Witten \cite{cates et witten}. They showed that the moments
of the harmonic measure, i.e., the Laplacian diffusion field near an
absorber, the latter taken as a simple random walk (RW, i.e., Brownian
motion), or self-avoiding walk (SAW, i.e., polymer), exhibit in $d$
dimensions multifractal scaling for $d<4.$ The associated exponents can be
recast as those of star copolymers made of a bunch of independent RW's
diffusing away from a generic point of the absorber. This allowed the
perturbative calculation of the Mf spectrum through standard RG theory for
polymers and the $\varepsilon =4-d$ expansion.

Star polymers or networks made of self-avoiding walks {\it only} and their
scaling properties are well understood \cite{duplantier4}, especially in 2D,
where all topology dependent exponents are known exactly from conformal
invariance \cite{duplantier4,duplantier2}. For simple random walks, scaling
exponents describing RW non-intersections \cite{duplantier5,lawler2},
conjectured in \cite{duplantier6}, have been recently derived using
conformal invariance and the so-called quantum gravity method \cite
{duplantier7}.

The aim of this Letter is to extend conformal invariance to arbitrary mixed
copolymers, thereby providing the exact Mf harmonic spectrum of a RW or a
SAW, an example of {\it conformal multifractality.}

Consider a general star copolymer ${\cal S}$ in the plane ${\Bbb R}^{2}$ (or
in ${\Bbb Z}^{2}$), made of an arbitrary mixture of Brownian paths or RW's $%
\left( \text{set}{\rm \;}{\cal B}\right) ,$ and polymers or SAW's $\left( 
\text{set}{\rm \;}{\cal P}\right) ,$ all starting at neighboring points. Any
pair $\left( A,B\right) $ of such paths, $A,B\in {\cal B}$ or ${\cal P},$
can be constrained in a specific way: either they avoid each other $\left(
A\cap B=\emptyset ,\text{ noted }A\wedge B\right) ,$ or they are transparent
and can cross each other (noted $A\vee B),$ corresponding to four different
fixed points \cite{joanny}. This notation allows for any {\it nested}
interaction structure; one can decide for instance that the branches $%
\left\{ P_{\ell }\in {\cal P}\right\} _{\ell =1,...,L}$ of an $L$-star
polymer, all mutually avoiding, further avoid a bunch of Brownian paths $%
\left\{ B_{k}\in {\cal B}\right\} _{k=1,...,n},$ all transparent to each
other: 
\begin{equation}
{\cal S}=\left( \bigwedge\nolimits_{\ell =1}^{L}P_{\ell }\right) \wedge
\left( \bigvee\nolimits_{k=1}^{n}B_{k}\right) .  \label{vw}
\end{equation}
In 2D the order of the branches of the star copolymer {\it %
does} matter and is intrinsic to our $\left( \wedge ,\vee \right) $ notation.

To each {\it specific} star copolymer center ${\cal S}$ is attached a
conformal scaling operator with a scaling dimension $x\left( {\cal S}\right)
.$ To obtain proper scaling we consider the Brownian paths and the polymers
to have the same mean size $R$. It is convenient to define for each star $%
{\cal S}$ a grand canonical partition function \cite
{duplantier4,duplantier5,ferber}, with fugacities $z$ and $z^{\prime }$ for
the total lengths $\left| {\cal B}\right| $ and $\left| {\cal P}\right| $ of
Brownian or polymer-like paths: 
\begin{equation}
{\cal Z}_{R}\left( {\cal S}\right) =\sum_{{\cal B},{\cal P}\in {\cal S}%
}z^{\left| {\cal B}\right| }z^{\prime \left| {\cal P}\right| }\;1\!{\rm I}%
_{R}\left( {\cal S}\right) ,  \label{zr}
\end{equation}
where the set of walks of ${\cal S}$ is constrained by the indicatrix$%
\;1\!{\rm I}_{R}\left( {\cal S}\right) $ to stay within a disc of radius $R$
centered on the star. At the critical values $z_{c}=\mu
_{B}^{-1},z_{c}^{\prime }=\mu _{P}^{-1},$ where for the RW's $\mu _{B}$ is
the coordination number of the underlying lattice, and $\mu _{P}$ the
effective one for the SAW's, ${\cal Z}_{R}$ decays as \cite
{duplantier4,ferber} 
\begin{equation}
{\cal Z}_{R}\left( {\cal S}\right) \sim R^{-x\left( {\cal S}\right) -x^{\vee
}},  \label{zrs}
\end{equation}
where $x\left( {\cal S}\right) $ is associated only with the singularity
occurring at the center of the star where all critical paths meet, while $%
x^{\vee }$ is the contribution of the independent dangling ends or split
star. It reads $x^{\vee }=\left\| {\cal B}\right\| x_{B,1}+\left\| {\cal P}%
\right\| x_{P,1}-2{\cal V},$ where $\left\| {\cal B}\right\| $ and $\left\| 
{\cal P}\right\| $ are respectively the total numbers of Brownian or polymer
paths of the star; $x_{B,1}$ or $x_{P,1}$ are the scaling dimensions of the
extremities of a {\it single} RW $x_{B,1}=0$, or SAW $x_{P,1}=\frac{5}{48}$%
\cite{duplantier4,nienhuis}. The last term ${\cal V}=\left\| {\cal B}%
\right\| +\left\| {\cal P}\right\| $ corresponds in (\ref{zrs}) to the
integration over extremity positions in the disc of radius $R$.

When the star is constrained to stay in a {\it half-plane} with its core
placed near the {\it boundary}, its partition function scales as \cite
{duplantier4,duplantier2} 
\begin{equation}
\tilde{{\cal Z}}_{R}\left( {\cal S}\right) \sim R^{-\tilde{x}\left( {\cal S}%
\right) -x^{\vee }},  \label{zz}
\end{equation}
where $\tilde{x}\left( {\cal S}\right) $ is the boundary scaling dimension, $%
x^{\vee }$ staying the same for star extremities in the bulk.

Any scaling dimension $x$ in the bulk is twice the {\it conformal dimension}
(c.d.) $\Delta ^{(0)}$ of the corresponding operator, while near a boundary
(b.c.d.) they are identical: 
\begin{equation}
x=2\Delta ^{\left( 0\right) },\quad \tilde{x}=\tilde{\Delta}^{\left(
0\right) }.  \label{xdelta}
\end{equation}
This Letter provides the main lines of a derivation of these exponents.

The idea is to use another representation where the RW's or SAW's are on a 2D
random lattice, i.e., in the presence of 2D {\it quantum gravity} \cite
{polyakov}. One can indeed put any 2D statistical system on a random planar
graph, thereby obtaining a new critical behavior, corresponding to the
confluence of the critical point of the infinite random graph with that of
the original model. The partition function of the random graph $G$ made of,
e.g., trivalent vertices, reads 
\begin{equation}
Z_{\chi }\left( \beta \right) =\sum\nolimits_{G(\chi )}e^{-\beta \left|
G\right| },  \label{zx}
\end{equation}
where the sum extends over graphs $G$ with a given topology of Euler
characteristic $\chi ,$ modulo the group of automorphisms of $G,$ $\left|
G\right| $ being the number of vertices. Near the critical point $\beta _{c}$
where $\left| G\right| $ becomes infinite 
\begin{equation}
Z_{\chi }\left( \beta \right) \sim \left( \beta -\beta _{c}\right)
^{2-\gamma _{{\rm str}}\left( \chi \right) };  \label{zxx}
\end{equation}
$\gamma _{{\rm str}}\left( \chi \right) =2-\frac{5}{4}\chi $ is the string
susceptibility exponent \cite{kostov}. The partition function of the
copolymer star ${\cal S}$ on the random lattice $G,$ with the sphere
topology $\left( \chi =2\right) ,$ is defined as:

\begin{equation}
Z\left( {\cal S}\right) =\sum\nolimits_{G\left( \chi =2\right) }e^{-\beta
\left| G\right| }{\cal Z}_{G}\left( {\cal S}\right) ,  \label{zs}
\end{equation}
where the partition function ${\cal Z}_{G}\left( {\cal S}\right) $ is
defined as in (\ref{zr}), with the indicatrix $1\!{\rm I}_{R}$ for the star
being confined to the disc of radius $R$ now being replaced by the
indicatrix $1\!{\rm I}_{G}$ of the star being embedded in $G.$ The partition
sum (\ref{zs}) is now three-fold grand canonical, depending implicitly on
fugacities $e^{-\beta },z,$ and $z^{\prime }.$ A further fugacity term has
to be added when dealing with a random graph $G$ with the {\it disc }%
topology $\left( \chi =1\right) $ and boundary length $\left| \partial
G\right| ,$ in order to define a boundary partition function

\begin{equation}
\tilde{Z}\left( {\cal S}\right) =\sum\nolimits_{G\left( \chi =1\right)
}e^{-\beta \left| G\right| -\beta ^{\prime }\left| \partial G\right| }\tilde{%
{\cal Z}}_{G}\left( {\cal S}\right) ,  \label{zzs}
\end{equation}
where the core star is now {\it on} the boundary ${\partial G}.$

There exists a finite size scaling regime \cite{duplantier3} where both the
lattice and the walks become infinite, $\beta ,\beta ^{\prime },z,$ and $%
z^{\prime }$ approaching together their respective critical values in a well
defined way. In this regime, the partition functions $Z$, ${\tilde{Z}}$,
after normalization by the random surface ones (\ref{zx}), are expected to
scale as \cite{duplantier8}: 
\begin{eqnarray}
Z\left( {\cal S}\right) /Z_{\chi =2}\left( \beta \right) &\sim &\left|
G\right| ^{-\Delta \left( {\cal S}\right) -\Delta ^{\vee }},  \label{zszx} \\
\tilde{Z}\left( {\cal S}\right) /Z_{\chi =1}\left( \beta \right) &\sim
&\left| \partial G\right| ^{-\tilde{\Delta}\left( {\cal S}\right) }\left|
G\right| ^{-\Delta ^{\vee }}.  \label{zzzx}
\end{eqnarray}

Here $\left| G\right| \sim \left( \beta -\beta _{c}\right) ^{-1}$ is the
average size of the random lattice, while $\left| \partial G\right| \sim
\left| G\right| ^{1/2}$ is the mean length of the boundary. $\Delta \left( 
{\cal S}\right) $ and $\tilde{\Delta}\left( {\cal S}\right) $ are
respectively the bulk and boundary conformal dimensions of the star core,
dressed by gravity. Finally the dimension $\Delta ^{\vee }$ in (\ref{zszx},%
\ref{zzzx}) is associated with the star extremities, as was $x^{\vee }$ in (%
\ref{zrs}). Equations (\ref{zszx}) and (\ref{zzzx}) are formally identical
to (\ref{zrs}) and (\ref{zz}) in the plane, after recalling (\ref{xdelta})
and identifying the star area $R^{2}$ in the plane to the random area $%
\left| G\right| $.

A general constitutive relation due to Knizhnik et al. exists between the
conformal dimension $\Delta ^{\left( 0\right) }$ of a scaling operator in
the {\it plane} and the c.d. $\Delta $ of the same operator on the {\it %
random surface} $\Delta ^{\left( 0\right) }=\Delta \left[ 1-\left( 1-\Delta
\right) /\kappa \right] ,$ where $\kappa $ is a parameter related to the
central charge of the original statistical model in the plane $c=1-6\left(
1-\kappa \right) ^{2}/\kappa $ \cite{polyakov,david}. For walks $c$
vanishes, whence $\kappa =3/2,$ and 
\begin{equation}
\Delta ^{\left( 0\right) }={U}\left( \Delta \right) ,\tilde{\Delta}^{\left(
0\right) }={U}\left( \tilde{\Delta}\right) ,{U}\left( x\right) =\frac{x}{3}%
\left( 1+2x\right) ,  \label{u}
\end{equation}
this relation holding also for {\it boundary} operators.

Here I give a set of basic underlying topological ``surgery'' rules which
allow the mixing of geometrical operators on a random surface.\ The set of
relations for walks is so stringent that it yields immediately the conformal
dimensions of any copolymer star, both on a random surface and in the plane.

{\it Star algebra}: The bulk and boundary conformal dimensions, in the
presence of gravity, satisfy: 
\begin{equation}
2\Delta -\gamma _{{\rm str}}\left( \chi =2\right) =\tilde{\Delta}.
\label{deltagamma}
\end{equation}
This fairly general relation can be derived from factorization properties of
partition functions like (\ref{zszx},\ref{zzzx}) \cite
{duplantier7,duplantier8}. As a consequence, substituting relation (\ref
{deltagamma}) with $\gamma _{{\rm str}}\left( \chi =2\right) =-\frac{1}{2}$
in (\ref{u}) gives the planar {\it bulk} dimensions from the gravity {\it %
boundary} ones: 
\begin{equation}
\Delta ^{\left( 0\right) }={U}\left( \Delta \right) ={V}\left( \tilde{\Delta}%
\right) ,\;{V}\left( x\right) =\frac{1}{24}\left( 4x^{2}-1\right) .
\label{v}
\end{equation}
Consider now two sub-stars $A,B,$ issued at the same point near a {\it %
boundary} line and avoiding each other. We state that their {\it boundary}
dimensions $\tilde{\Delta}$ on the{\it \ random surface} are {\it additive}: 
\begin{equation}
\tilde{\Delta}\left( A\wedge B\right) =\tilde{\Delta}\left( A\right) +\tilde{%
\Delta}\left( B\right) .  \label{wedge}
\end{equation}
Owing to (\ref{deltagamma}), a similar relation, but with a constant shift,
exists for bulk exponents. On $G,$ two elements in a pair $\left( A,B\right) 
$ of walk sets are indeed made non-intersecting by gluing a fluctuating
patch of random surface between them.\ Each set, $A$ or $B,$ defines its own
independent disc as in (\ref{zzs}), with additive b.c.d.'s (\ref{zzzx},\ref
{wedge}). So, {\it mutually avoiding sets are rendered independent by the
fluctuations of a random surface.} Thus relation (\ref{wedge}) can be
derived from factorization properties of exact partition functions \cite
{duplantier7,duplantier8}.

Consider by contrast two sets $A,B$ of walks which are mutually transparent,
i.e., $A\vee B.$ In the half-plane ${\Bbb R}^{+}\times {\Bbb R},$ they are
independent, and their boundary dimensions obey a linear relation 
\begin{equation}
\tilde{\Delta}^{\left( 0\right) }\left( A\vee B\right) =\tilde{\Delta}%
^{\left( 0\right) }\left( A\right) +\tilde{\Delta}^{\left( 0\right) }\left(
B\right) ,  \label{wedge0}
\end{equation}
due to the trivial factorization of their partition functions. On a random
surface, their boundary dimensions are obtained by inverting (\ref{u}) 
\begin{equation}
\tilde{\Delta}=U^{-1}\left( \tilde{\Delta}^{\left( 0\right) }\right)
,\;U^{-1}\left( x\right) =\frac{1}{4}\left( \sqrt{24x+1}-1\right) ,
\label{u-1}
\end{equation}
and are not additive. The metric fluctuations indeed {\it couple} the sets $%
A,B.$

It is clear at this stage that the set of equations above is {\it complete.}
It allows for the calculation of any conformal dimensions $\Delta \left( 
{\cal S}\right) $ or $\Delta ^{\left( 0\right) }\left( {\cal S}\right) $
associated with a star structure ${\cal S}$ of the most general type, as in (%
\ref{vw}), involving $\left( \wedge ,\vee \right) $ operations separated by
nested{\it \ }parentheses. Any such structure can be systematically reduced:
one starts from the outermost parenthesis set, and calculates b.c.d.'s of
operations $\left( \wedge ,\vee \right) $ by using (\ref{wedge}) for $\wedge 
$ on the random surface $G,$ and (\ref{wedge0}) for $\vee $ on the plane $%
{\Bbb R}^{2},$ while applying repeatedly the non-linear map $U:G\rightarrow 
{\Bbb R}^{2}$ (\ref{u}), or its inverse $U^{-1}$ (\ref{u-1}) to transfer to
the proper space where the boundary dimension is a linear representation of $%
\wedge $ or $\vee .$ At the end, one uses (\ref{v}) to recover bulk
dimensions.

{\it Brownian-polymer exponents: }The single extremity scaling dimensions
are for a RW or a SAW near a Dirichlet boundary in ${\Bbb R}^{2}$ \cite
{duplantier4,cardy} 
\begin{equation}
\tilde{\Delta}_{B}^{\left( 0\right) }\left( 1\right) =1,\;\tilde{\Delta}%
_{P}^{\left( 0\right) }\left( 1\right) =%
%TCIMACRO{\tfrac{5}{8} }
%BeginExpansion
{\textstyle {5 \over 8}}%
%EndExpansion
,  \label{num}
\end{equation}
or on $G,$ using (\ref{u-1}), $\tilde{\Delta}_{B}\left( 1\right)
=U^{-1}\left( 1\right) =1,\;\tilde{\Delta}_{P}\left( 1\right) =U^{-1}\left( 
%TCIMACRO{\tfrac{5}{8}}
%BeginExpansion
{\textstyle {5 \over 8}}%
%EndExpansion
\right) =%
%TCIMACRO{\tfrac{3}{4}}
%BeginExpansion
{\textstyle {3 \over 4}}%
%EndExpansion
.$ Because of the star algebra described above these are the only numerical
seeds, i.e., generators, we need.

Stars can include bunches of $n$ copies of transparent RW's or $m$
transparent SAW's. Their b.c.d.'s in ${\Bbb R}^{2}$ are respectively, by
using (\ref{wedge0}) and (\ref{num}), $\tilde{\Delta}_{B}^{\left( 0\right)
}\left( n\right) =n$ and $\tilde{\Delta}_{P}^{\left( 0\right) }\left(
m\right) =\frac{5}{8}m,$ from which the inverse mapping (\ref{u-1}) to the
random surface yields $\tilde{\Delta}_{B}\left( n\right) =U^{-1}\left(
n\right) $ and $\tilde{\Delta}_{P}\left( m\right) =U^{-1}\left( 
%TCIMACRO{\tfrac{5}{8}}
%BeginExpansion
{\textstyle {5 \over 8}}%
%EndExpansion
m\right) .$ The star made of $L$ bunches $\ell \in \left\{ 1,...,L\right\} $
of $n_{\ell }$ transparent RW's each, and $L^{\prime }$ bunches $\ell
^{\prime }\in \left\{ 1,...,L^{\prime }\right\} $ of $m_{\ell ^{\prime }}$
SAW's, all mutually avoiding, has planar scaling dimensions owing to (\ref{u}%
,\ref{v}) and (\ref{wedge}) 
\begin{eqnarray*}
\tilde{\Delta}^{\left( 0\right) }\left\{ n_{\ell },m_{\ell ^{\prime
}}\right\} &=&U\left( \tilde{\Delta}\right) ,\;\Delta ^{\left( 0\right)
}\left\{ n_{\ell },m_{\ell ^{\prime }}\right\} =V\left( \tilde{\Delta}%
\right) , \\
\tilde{\Delta}\left\{ n_{\ell },m_{\ell ^{\prime }}\right\}
&=&\sum\nolimits_{\ell =1}^{L}U^{-1}\left( n_{\ell }\right)
+\sum\nolimits_{\ell ^{\prime }=1}^{L^{\prime }}U^{-1}\left( 
%TCIMACRO{\tfrac{5}{8} }
%BeginExpansion
{\textstyle {5 \over 8}}%
%EndExpansion
m_{\ell ^{\prime }}\right) .
\end{eqnarray*}
These exponents are {\it invariant }under {\it permutation }of the bunches
of walks. The existence of such a relation has been found for RW's in \cite
{lawler}, but with an unspecified $U,$ which is here derived from quantum
gravity and generalized to SAW's.

For a copolymer star ${\cal S}_{L,L^{\prime }}$ made of $L$ RW's and $%
L^{\prime }$ SAW's, all mutually avoiding $\left( \forall \ell ,\ell
^{\prime },n_{\ell }=m_{\ell ^{\prime }}=1\right) ,$\ $\tilde{\Delta}\left( 
{\cal S}_{L,L^{\prime }}\right) =L+\frac{3}{4}L^{\prime }$ gives the c.d. in 
${\Bbb R}^{2}$%
\begin{eqnarray*}
\tilde{\Delta}^{\left( 0\right) }\left( {\cal S}_{L,L^{\prime }}\right)  &=&%
%TCIMACRO{\tfrac{1}{3}}
%BeginExpansion
{\textstyle {1 \over 3}}%
%EndExpansion
\left( L+%
%TCIMACRO{\tfrac{3}{4}}
%BeginExpansion
{\textstyle {3 \over 4}}%
%EndExpansion
L^{\prime }\right) \left( 1+2L+%
%TCIMACRO{\tfrac{3}{2}}
%BeginExpansion
{\textstyle {3 \over 2}}%
%EndExpansion
L^{\prime }\right)  \\
\Delta ^{\left( 0\right) }\left( {\cal S}_{L,L^{\prime }}\right)  &=&%
%TCIMACRO{\tfrac{1}{24}}
%BeginExpansion
{\textstyle {1 \over 24}}%
%EndExpansion
\left[ 4\left( L+%
%TCIMACRO{\tfrac{3}{4}}
%BeginExpansion
{\textstyle {3 \over 4}}%
%EndExpansion
L^{\prime }\right) ^{2}-1\right] , 
\end{eqnarray*}
recovering for $L=0$ the SAW exponents \cite{duplantier2} and for $L^{\prime
}=0$ the RW non-intersection exponents \cite{duplantier7}.

{\it Disconnection exponents: }Take any walk star $A,$ with elementary
boundary c.d. $\tilde{x}$ in ${\Bbb R}^{2}.$ The star ${\cal S}_{n}=\left( \vee
A\right) ^{n}$ made of $n$ transparent copies of $A$ has b.c.d. $n\tilde{x}$
in ${\Bbb R}^{2},$ thus $U^{-1}\left( n\tilde{x}\right) $ on $G.$ Its bulk
c.d. in ${\Bbb R}^{2}$ is, according to (\ref{xdelta}, \ref{v}), $x\left( 
{\cal S}_{n}\right) =2V\left[ U^{-1}\left( n\tilde{x}\right) \right] ,$
which differs from $nx,$ where $x=2V\left[ U^{-1}\left( \tilde{x}\right)
\right] $ is the bulk c.d. in ${\Bbb R}^{2}.$ Their difference is the {\it %
disconnection} exponent, governing the conditioned probability $P_{R}\sim
R^{-x\left( {\cal S}_{n}\right) +nx}$ that the union of $n$ copies does not
disconnect the star origin from infinity, within a disc of radius $R.$

{\it Multifractal harmonic measure: }The harmonic measure $H\left( w\right) $
of a given set is the probability that a RW coming from infinity first hits
the set (the absorber) at point $w.$ When the set is a RW or a SAW of size $%
R,$ the site average of its moments $H^{n}$ has been shown \cite{cates et
witten} to be represented by a copolymer star partition function of type (%
\ref{vw}) where the absorber avoids a bunch of $n$ independent RW's.\ More
precisely $\sum\nolimits_{w}H^{n}\left( w\right) \sim {\cal Z}_{R}\left( 
{\cal S}_{\wedge }n\right) /{\cal Z}_{R}\left( {\cal S}_{\wedge }1\right) ,$
where the absorber ${\cal S}$ is either the two-RW star $B\vee B$ or the
two-SAW star $P\wedge P,$ made of two non-intersecting SAW's.\ We have
introduced the short-hand notation ${\cal S}_{\wedge }n\equiv {\cal S}\wedge
\left( \vee B\right) ^{n}$ describing the copolymer star made by the
absorber ${\cal S}$ hit by the bunch $\left( \vee B\right) ^{n}$ at the apex
only.\ Owing to Eq.(\ref{zrs}), we get the scaling $\sum\nolimits_{w}H^{n}%
\left( w\right) \sim R^{-\tau \left( n\right) },$ where $\tau \left(
n\right) =\left( n-1\right) D\left( n\right) =x\left( {\cal S}_{\wedge
}n\right) -x\left( {\cal S}_{\wedge }1\right) $ defines (annealed \cite
{halsey3}) generalized dimensions $D\left( n\right) .$ Our formalism (\ref{v}%
,\ref{wedge},\ref{u-1}) immediately gives the scaling dimensions $x\left( 
{\cal S}_{\wedge }n\right) =2V\left( \tilde{\Delta}\left( {\cal S}\right)
+U^{-1}\left( n\right) \right) ,$ where $\tilde{\Delta}\left( {\cal S}%
\right) $ is as usual the quantum gravity boundary dimension of the absorber 
${\cal S}$ alone. A simple calculation gives $\tau \left( n\right) $ and its
Legendre transform $f\left( \alpha \right) +\tau \left( n\right) =\alpha
n,\;\alpha =d\tau \left( n\right) /dn$%
\begin{eqnarray}
\tau \left( n\right) &=&\frac{1}{2}\left( n-1\right) +y\frac{1}{24}\left( 
\sqrt{24n+1}-5\right) , \\
f\left( \alpha \right) &=&\frac{1}{24}\left\{ \frac{25}{2}+5y-\frac{1}{2}%
y^{2}\left( 2\alpha -1\right) ^{-1}-\alpha \right\} , \label{mf}
\end{eqnarray}
where $y\equiv 4\tilde{\Delta}\left( {\cal S}\right) -1$ is the only
parameter encoding which absorber we consider (which can actually be any
star tip). For a RW absorber, we have $\tilde{\Delta}\left( B\vee B\right)
=U^{-1}\left( 2\right) =\frac{3}{2},$ while for a SAW $\tilde{\Delta}\left(
P\wedge P\right) =2\tilde{\Delta}_{P,1}=2U^{-1}\left( \frac{5}{8}\right) =%
\frac{3}{2},$ thus $y=5$ in both cases. The coincidence of these two values
tells us that in 2D the harmonic multifractal spectra $f\left( \alpha
\right) $ of a random walk or a self-avoiding walk are {\it identical.}
Their Mf spectra associated with walk {\it ends} \cite{cates et witten}
however differ{\it ,} and are obtained using $y=3$ for a RW end, or $y=2$ for
a SAW end. 
\begin{figure}
%\epsfxsize=6truecm{\centerline{\epsfbox{fg1.eps}}}
%\vskip -7pt
\centerline{\epsfig{file=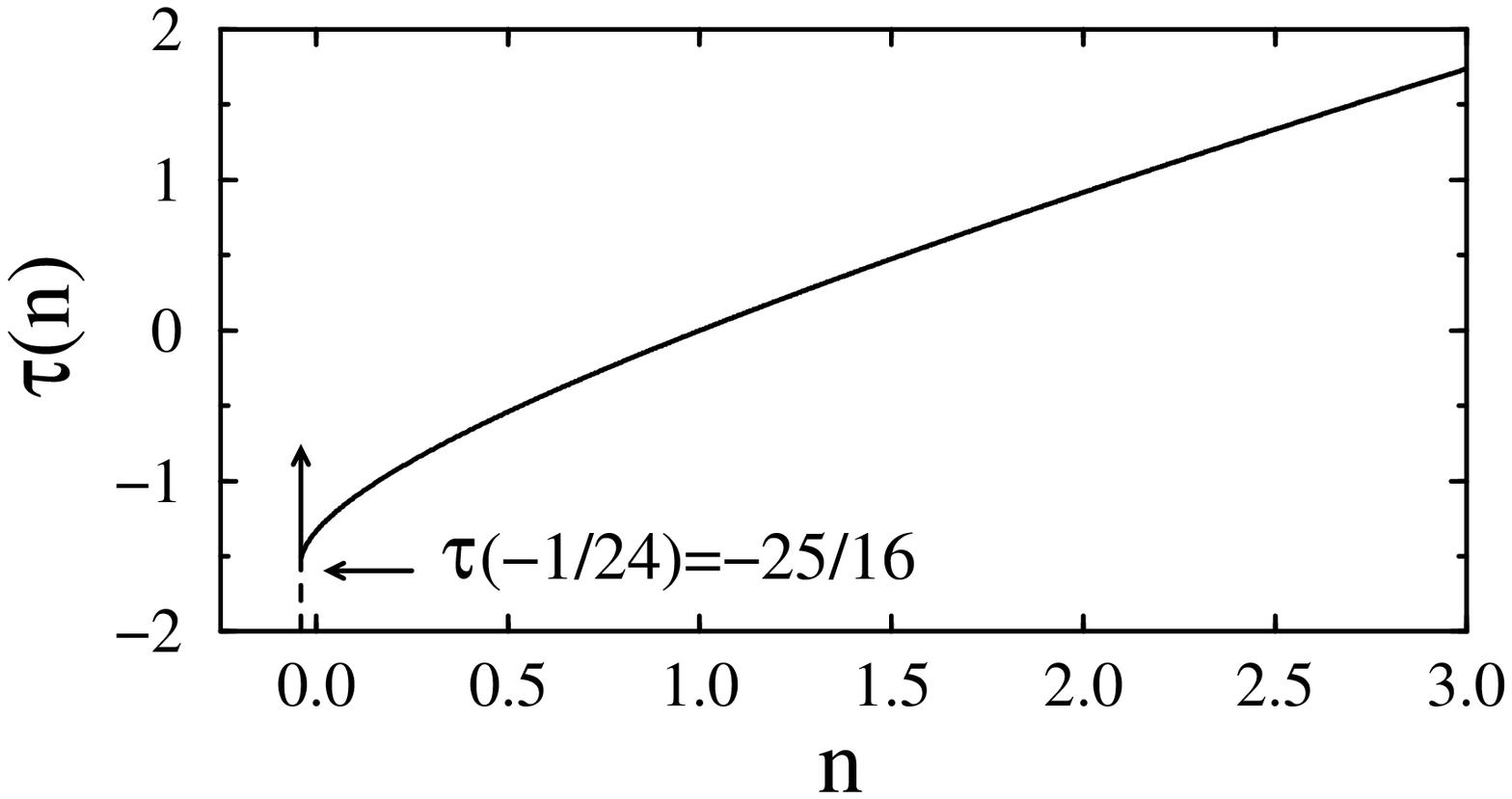,width=8.5cm}}
%\medskip
%\label{Figure}
\end{figure}
\begin{figure}
\vskip -23pt
\centerline{\epsfig{file=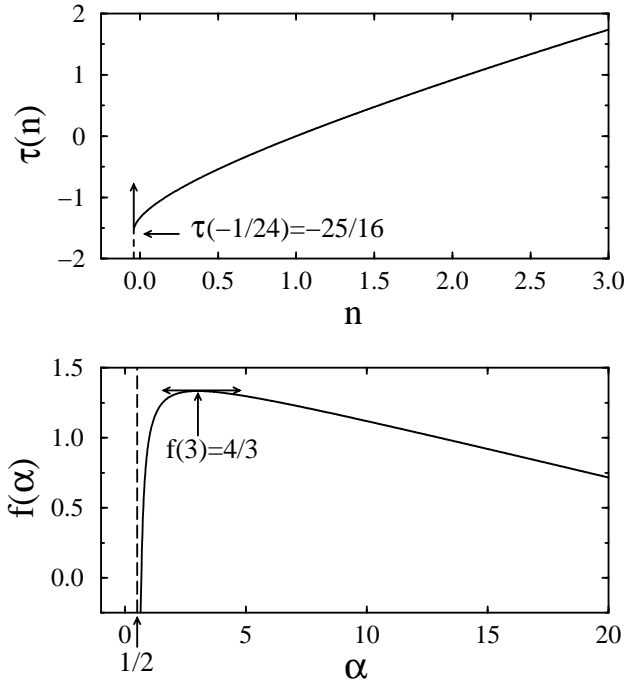,width=8.5cm}}
\smallskip
\caption{Harmonic multifractal dimensions $\tau (n)$ and spectrum
$f(\alpha)$ of a two-dimensional RW or SAW.}
\label{Figure}
\end{figure}
The corresponding universal curves for $y=5$ are shown in Fig.1: $\tau \left( 
n\right) 
$ is half a parabola, and $f\left( \alpha \right) $ a hyperbola. $D\left(
1\right) =\tau ^{\prime }\left( 1\right) =1$ is just Makarov's theorem \cite
{makarov}; the divergence of $f$ at $\alpha _{\min }=\frac{1}{2}$
corresponds to singular needles in the absorber, while $-\tau \left(
0\right) =\sup_{\alpha }f\left( \alpha \right) =f\left( 3 \right) =\frac{4}{3}$ 
is the
Hausdorff dimension of the Brownian frontier or of a SAW. Thus Mandelbrot's
classical conjecture identifying the latter two is generalized and proven
for the whole $f\left( \alpha \right) $ harmonic spectrum.

{\bf Acknowledgements.} I thank T.C. Halsey for discussions and a careful
reading of the manuscript.

\end{document}